\documentclass[aps,prd,twocolumn,superscriptaddress,showpacs]{revtex4-1}
\usepackage[utf8]{inputenc}
\usepackage{amsmath}
\usepackage{amsfonts}
\usepackage{amssymb}
\usepackage{graphics}
\usepackage{tabularx}
\usepackage{mathtools}
\usepackage{float}
\begin{document}
\title{Semi-phenomenological analysis of neutron scattering results for quasi-two dimensional quantum anti-ferromagnet}
\author{Subhajit Sarkar}
\email{subhajit@bose.res.in}
\author{Ranjan Chaudhury}
\email{ranjan@bose.res.in}
\author{Samir K. Paul}
\email{smr@bose.res.in}
\affiliation{S. N. Bose National Centre For Basic Sciences, Block - JD, Sector - III, Salt Lake, Kolkata - 700098, India}
\date{\today}
\begin{abstract}\label{abs}
The available results from the inelastic neutron scattering experiment performed on the quasi-two dimensional spin $\frac{1}{2}$ anti-ferromagnetic material $La_2 Cu O_4$ have been analysed theoretically. The formalism of ours is based on a semi-classical like treatment involving a model of an ideal gas of mobile vortices and anti-vortices built on the background of the N$\acute{e}$el state, using the bipartite classical spin configuration corresponding to an XY- anisotropic Heisenberg anti-ferromagnet on a square lattice. The results for the integrated intensities for our spin $\frac{1}{2}$ model corresponding to different temperatures, show occurrence of vigorous unphysical oscillations, when convoluted with a realistic spectral window function. These results indicate failure of the conventional semi-classical theoretical model of ideal vortex/anti-vortex gas arising in the Berezinskii-Kosterlitz-Thouless theory for the low spin magnetic systems. A full fledged quantum mechanical formalism and calculations seem crucial for the understanding of topological excitations in such low spin systems. Furthermore, a severe disagreement is found to occur at finite values of energy transfer between the integrated intensities obtained theoretically from the conventional formalism and those obtained experimentally. This further suggests strongly that the full quantum treatment should also incorporate the interaction between the fragile-magnons and the topological excitations. This is quite plausible in view of the recent work establishing such a process in XXZ quantum ferromagnet on 2D lattice. The high spin XXZ quasi-two dimensional antiferromagnet like $MnPS_3$ however follows the conventional theory quite well.

\textit{Highlights:}
\begin{itemize}
\item Inadequacies in the conventional meron gas phenomenology in explaining the spin dynamics \\ corresponding to layered low-spin anti-ferromagnets.
\item Requirement of a full fledged quantum mechanical formalism for the understanding of spin \\ dynamics induced by both topological and conventional excitations. 
\item Necessity of a proper understanding of the interaction between the conventional and \\ topological excitations at the quantum level.
\end{itemize}
\end{abstract}
\maketitle
\section{Introduction}
Spin dynamics in low dimensional magnetic systems have generated a significant research interest during the last three decades\cite{Kat,Pro,Cwi,TC,Cha,Web,Yus,Ulb,Lew,Bra,Ran2011,Yon,Kje,Hei,Sti,Lov,Rei,Bet,End,Yos,Hir,Hut,Day,Ale,Wie,Ste,Cap,Cas,Yam, Yak, Yen}. Different types of one and quasi-one dimensional as well as two and quasi-two dimensional systems have been studied both experimentally and theoretically to probe and understand the spin dynamics arising from the conventional spin wave excitations and their mutual interactions, as well as from vortex/meron and soliton  like topological excitations \cite{Lov, Hei, Kje, Ste, Yak, Rei}. In many of these systems the topological excitations of soliton and vortex/meron type occur naturally 
as they are thermodynamically feasible.
\paragraph*{}
Motivated by the distinct possibilities of applications towards building of magnetic devices, the quasi-two dimensional systems have attracted a renewed research interest in recent times. Magnetic vortices present in these systems, have proved to be potential candidates for switching devices \cite{Ber, Wac, Cow, Van}. Direct experimental evidences of such vortices have been verified by both the Magnetic Force Microscopy (MFM) and the spin-polarized Scanning Tunnelling Microscopy (STM) \cite{Hau, West}.
\paragraph*{}
The spin dynamics in many of the above magnetic systems has been investigated experimentally using the Inelastic Neutron Scattering (INS) and  the Nuclear Magnetic Resonance (NMR) techniques. These include quasi-one dimensional systems such as $CsNiF_3$,  layered systems such as $K_2CuF_4$, $Rb_2 CrCl_4$, $LiCrO_2$, magnetically intercalated graphites, layered ruthenates, layered manganites and the high-$T_C$ cuprates \cite{Hei,Kje, Sti, Rei, Yos, Hir, Hut, Day, Ale, Wie, Ste, Cap, Cas, Yam}. In the INS experiments performed on several of  above materials, the existence of a prominent ``central peak" (at $
\hbar \omega = 0$) has been confirmed in the plot for the dynamical structure function $S(\mathbf{q}, \omega)$ vs. neutron energy transfer `$\hbar\omega $' in the constant `$\mathbf{q}$’ scan \cite{Hir, Yos}. These findings further serve as the motivation behind the huge variety of experiments performed on the layered magnetic systems. Moreover, the advancement of numerical and computational techniques also contributed to the understanding of the possible role of both the spin waves and the topological excitations in emergence of the central peak.
\paragraph*{}
Kosterlitz and Thouless, and Berezniskii independently introduced the concept of vortex and anti-vortex like topological spin excitations in the two dimensional classical magnetic (spin) systems \cite{KT, Bez, Jose}. According to their ideas, there exists a non-conventional topological phase transition (known as Berezinskii- Kosterlitz- Thouless or BKT transition) characterized by the crossover between binding to unbinding phases of vortex- anti-vortex pairs at a transition temperature $T_{BKT}$. Below this temperature all the vortices and anti-vortices are in a bound state and above this temperature some of them start moving freely. Further analytical and numerical studies and suitable extension of these approaches led to the proposal for the existence of topological vortices and anti-vortices in pure XY model and merons and anti-merons in XY- anisotropic Heisenberg model, for both ferromagnetic and anti-ferromagnetic types, on two- dimensional lattices \cite{Hub, Mer, Gou, Vol, Wys, And}. Furthermore, approximate analytical calculations and Monte Carlo-Molecular Dynamics (MCMD) simulations have strongly suggested that the freely moving topological excitations in the regime $T > T_{BKT}$, contribute non-trivially to the spin-spin correlation and give rise to the ``central peak", as mentioned above \cite{Hub, Mer, Gou, Vol}. The occurrence of such a ``central peak" in quasi-two dimensional magnetic systems is now unanimously believed to be the signature for the dynamics of  mobile topological excitations in a layer.
\paragraph*{}
  In spite of a lot of studies on the anisotropic Heisenberg model on two dimensional lattices the contributions of vortices/anti-vortices to the dynamics of the model, especially the detailed quantitative features of the DSFs are still not completely understood.
\paragraph*{}
In the context of two-dimensional magnetism, the undoped (anti-ferromagnetic and non-superconducting) phases of the high $T_c$ cuprate systems are believed to be excellent examples of two dimensional XY anisotropic Heisenberg Hamiltonian in an appropriate temperature regime. One member of this class of systems is $La_2 CuO_4$, on which extensive INS experiments have been performed \cite{Yen, Yam}. This is a truly spin-1/2 layered anti-ferromagnet. The intra-layer integrated intensity corresponding to the results of INS experiment performed on $La_2 CuO_4$, exhibits a central peak when plotted against the neutron energy transfer $\hbar \omega$ (or frequency `$ \omega $').
\paragraph*{}
It has been shown that the results of vortex gas phenomenology and numerical simulations lead to an anomaly in the case of layered anti-ferromagnetic systems having very low spin values (S=1/2) \cite{RC}. Strikingly enough, the value of $T_{BKT}$ obtained from Renormalization group analysis and numerical simulations is four (4) times the value of $T_{BKT}$ calculated from the classical expression obtained by Kosterlitz and Thouless \cite{RC}. Furthermore, in our previous work we have already established that for quasi-two dimensional ferromagnetic systems having low spin values (S= 1/2) the conventional semi-classical like treatment involving the ideal gas of unbound vortices/merons and anti-vortices/anti-merons corresponding to high temperature regime $T>T_{BKT}$, shows large inconsistency with the experimental situation and exhibits unphysical behaviour \cite{Sar}. In  this case the theoretical dynamical structure function (DSF) turns out to be negative for a wide range of energy transfers! However the range, over which the theoretical DSF remains positive, increases when the value of the spin is increased \cite{Sar}. These facts motivate us to investigate and test in detail the applicability of the semi-classical-like treatment mentioned above to the INS results corresponding to real anti-ferromagnetic systems with S=1/2. For this exercise we select $La_2 CuO_4$ as the reference system \cite{Yen, Yam}.
\paragraph*{}
It is worthwhile  to mention that the BKT transition can also be identified at the transition temperature $T_{BKT}$ where the spin-stiffness jumps discontinuously from a universal value below $T_{BKT}$ to zero above $T_{BKT}$ \cite{Hara}. Moreover, in anti-ferromagnetic model possessing Ising like anisotropy (in the z-direction) on two dimensional lattices, with external field being applied in the z-direction, such a discontinuous jump has been observed \cite{Holt}.  However, the discontinuous jump in this case may have its origin different from the vortex-anti-vortex unbinding mechanism since it is well known that the BKT transition in magnetic systems can only occur when the anisotropy is XY like \cite{Hub, Mer, Gou, Vol}.
\paragraph*{}
The plan of the paper is as follows:- in section II we describe the formulation of our semi-classical like treatment; in section III we discuss our calculations and results and finally in section IV the conclusions and discussions of our present investigation are presented.
\section{Mathematical formulation}
The dynamics of mobile topological excitations in an anti-ferromagnetic system on a two-dimensional square lattice have been analysed both analytically and numerically \cite{Hub, Mer, Gou, Vol, Wys}. The analytical studies have been performed by assuming a classical ideal gas of vortices/merons where the vortices/merons obey Maxwell's velocity distribution. The model system is described  the XY-anisotropic Heisenberg (XXZ) Hamiltonian, viz.,
\begin{equation}\label{eq:1}
\mathcal{H} = -J \sum_{\langle ij,pq \rangle} (S^{x}_{ij}S^{x}_{pq} + S^{y}_{ij}S^{y}_{pq} + \lambda S^{z}_{ij}S^{z}_{pq}),
\end{equation}
where $\langle ij,pq \rangle$ label the nearest neighbour sites on a two-dimensional square lattice and $J (<0)$ is the anti-ferromagnetic exchange coupling. Here $\lambda$ is the anisotropy parameter whose pure XY and isotropic Heisenberg limit correspond to $\lambda=0$ and 1 respectively. 
\paragraph*{}
The structures of the vortices/merons have been obtained by solving the classical equations of motion corresponding to the Hamiltonian given by eqn. (\ref{eq:1}). In deriving the classical equations of motion the spins have been considered to be classical objects (classical spin fields $S(\mathbf{r}, t)$) as a function of position coordinates and time, which are defined on the entire lattice. At even or odd lattice sites these spin fields become identical to the following bi-partite spin configurations:
\begin{widetext}
\begin{eqnarray}\label{eq:2}
S_{ij}^{even}  &=& + S[\sin(\Theta_{ij} + \theta_{ij})\, \cos(\Phi_{ij} + \phi_{ij}),\,\sin(\Theta_{ij} + \theta_{ij})
\sin(\Phi_{ij} + \phi_{ij}),\, \cos(\Theta_{ij} + \theta_{ij}))] , \nonumber \\
S_{ij}^{odd}  &=& - S[\sin(\Theta_{ij} - \theta_{ij})\cos(\Phi_{ij} - \phi_{ij}),\, \sin(\Theta_{ij} - \theta_{ij})\,
\sin(\Phi_{ij} - \phi_{ij}),\, \cos(\Theta_{ij} - \theta_{ij}))],
\end{eqnarray}
\end{widetext}
where `$even$' and `$odd$' signifies the two different sub-lattices \cite{Mik}. The static spin configuration corresponding to the merons are described by the capital angles $\Theta(\mathbf{r}) $ (polar) and $\Phi(\mathbf{r})$ (azimuthal), and the time dependent small angles $\theta(\mathbf{r}, t)$ and $\phi(\mathbf{r}, t)$ describes the corresponding deviations from the static structure due to the motion of the merons and the spin dynamics above BKT transition temperature \cite{Vol, Wys}. The expression of the vortex core radius is given by \cite{Vol, Wys}
\begin{equation} \label{eq:r}
 r_v = \frac{a}{2} \sqrt{\frac{\lambda}{1-\lambda}}.
\end{equation}

\paragraph*{}
From the above considerations the in-plane dynamical structure function (in-plane DSF) $S^{xx} (\mathbf{q}, \omega$) is 
given by,
\begin{equation}\label{eq:3}
S^{xx} (\mathbf{q}, \omega) = \frac{S(S+1)}{2\pi} \frac{\gamma ^{3} \xi ^{2}}{(\omega ^{2} + \gamma ^{2}[1+ (\xi \mathbf{q}
^{*})^{2}])^2},
\end{equation}
with $\gamma = \frac{\sqrt{\pi}\bar{u}}{2\xi}$, where $\mathbf{q}^{*} = (\mathbf{q}_{0} - \mathbf{q})$; $\mathbf{q}_{0}= (\pi/a, \pi/a)$ and in our case $S=\frac{1}{2}$. The above expression for the in-plane dynamical structure function is a squared Lorentzian exhibiting a central peak at $\omega =0$ in `$\omega$'-space for constant $\mathbf{q}$- scan and exhibiting a central peak at the zone boundary of the first Brillouin Zone (BZ) in the `$q$' space for constant $\omega$-scan  \cite{Vol, Wys}. In the above expression $\bar{u}$ is the root mean square (rms) velocity of the vortices and is given by,
\begin{equation}\label{eq:4}
\bar{u}= \sqrt{b\pi}\frac{JS(S+1)a^2}{\hbar}(\sqrt{n^f_v}) \,\, \tau^{-1/4},
\end{equation}
where $n^f_v \sim (2\xi)^{-2}$ is the density of free vortices at $T>T_{BKT}$ \cite{Hub}. Here $\xi= \xi_0 e^{b/\sqrt{\tau}}$ is the intra-layer two-spin correlation length due to the presence of vortices, where $\xi_0$ is of the order of lattice spacing `$a$'; $\tau = (\frac{T}{T_{BKT}}-1)$ is the reduced temperature and $b$ is a dimensionless parameter whose numerical value is generally around $1.5$ \cite{RC, Hir} . The quantity $S^{xx} (\mathbf{q}, \omega)$ is sensitive to the in-plane structure of the vortices/merons  \cite{Vol, Wys}.
\paragraph*{}
Again the effective analytical expression for the out-of-plane dynamical structure function (out-of-plane DSF) $S^{zz} (\mathbf{q},\omega)$ in the limit of very small `q' is given by  \cite{Vol, Wys}
\begin{equation}\label{eq:5}
S^{zz} (\mathbf{q},\omega) = \frac{n_v^f \bar{u}}{32(1+\lambda)^2 J^2 \sqrt{\pi} q^3} exp[-(\frac{\omega}{\bar{u}q})^2]. 
\end{equation}
The above form of the out-of-plane dynamical structure function is a Gaussian, exhibiting again a central peak at $\omega = 0$, when plotted in the constant $\mathbf{q}$-scan. The function $S^{zz} (\mathbf{q},\omega)$ is sensitive to the out-of-plane shape of the vortices/merons \cite{Vol, Wys}.
\paragraph*{}
In the case of layered systems, in a suitable regime in the parameter space comprising of temperature and wave vector where these systems behave effectively as two-dimensional systems, the integrated intensity corresponding to a typical inelastic neutron scattering experiment is given by
\begin{equation}\label{eq:6}
I(\omega) = \int \sum_{\alpha} S^{\alpha \alpha}(\mathbf{q_{2D}}, \omega) dq_x \, dq_y,
\end{equation}
where the quantity $S^{\alpha \alpha}(\mathbf{q_{2D}}, \omega)$ represents the intra-layer in-plane spin dynamical structure function when $\alpha=x$ and $y$ and the intra-layer out-of-plane spin dynamical structure function when $\alpha = z$ \cite{Lov, Gen, Hip}.
\paragraph*{}
The approach in our present work is quite similar to that adopted earlier in the case of ferromagnetic systems \cite{Sar}. In order to compare theory with experiment the dynamical structure function, obtained from the model under consideration, is multiplied with the resolution function $R(t)$ (in time domain) or convoluted with $\tilde{R}(\omega - \omega^{\prime})$ (in the frequency domain), as has been done in the case of  ferromagnetic system \cite{Gen, Tak, Sar}. Hence, the components of the convoluted integrated intensity comes out to be,
\begin{equation}\label{eq:7}
I_{conv}^{\alpha \alpha}(\omega) =  \int  dq_x \, dq_y \int d\omega^{\prime} \tilde{R}(\omega - \omega^{\prime}) S^{\alpha 
\alpha}(\mathbf{q_{2D}}, \omega^{\prime}).
\end{equation}
The resolution function has to be chosen in such a way that minimum ripples occur at the end points of the resolution width. The different parameters of the resolution function can be obtained from the resolution half width or the full width at the half maximum (FWHM) which are quoted in the experiments. Since X and Y components of the spins are symmetric i.e., $S^{xx}(\mathbf{q},\, \omega)=S^{yy}(\mathbf{q},\, \omega)$ the total intensity comes out to be, 
\begin{equation}\label{eq:8}
I(\omega) = 2I^{xx}(\omega)+I^{zz}(\omega).
\end{equation}
\paragraph*{}
Furthermore, keeping in mind the low spin situation the quantum mechanical detailed balance condition is incorporated in our formalism \cite{Tol}. Then semi-classical estimate for $I(\omega)$, denoted by $I_{SC}(\omega)$ is recovered by the relation,
\begin{equation}\label{eq:9}
I^{SC} (\omega) = \frac{2}{1 + exp(\frac{-\hbar \omega}{k_B T})} I(\omega),
\end{equation} 
where the factor $\frac{2}{1 + exp(\frac{-\hbar \omega}{k_B T})}$ is the detailed balance factor and is called the Windsor 
factor \cite{Eva, Win}. The superscript `SC' stands for the term semi-classical.
\paragraph*{}
As has been pointed out in the case of ferromagnetic systems, this approach of ours is truly `semi-classical-like' in the sense that the expression for the rms vortex velocity `$\bar{u}$', as given in eqn. (4), contains `$\hbar$' and in addition the quantum mechanical detailed balance condition has been incorporated through the Windsor factor\cite{Sar}.
\paragraph*{}
It is worthwhile to mention that the above formulation based on dilute vortex/meron gas phenomenology hold for unbound  anti-vortices/anti-merons too on the basis of the assumption that the vortices/merons and anti-vortices/anti-merons do not interact with each other.
\section{Calculations and results}
In this work the formalism of Section II is applied to an anti-ferromagnetic material $La_2 CuO_4$ on which inelastic neutron scattering experiments (INS) involving polarized neutron beam have been performed \cite{Yen, Yam}. The material is an XY-anisotropic quasi-two-dimensional spin 1/2 quantum Heisenberg anti-ferromagnet. The magnetic lattice structure of it is composed of stacking of two-dimensional square lattices \cite{Yen, Yam}. The spin Hamiltonian relevant to the above material is given by,
\begin{equation}\label{eq:10}
\mathcal{H} = \underbrace{( -J\sum_{<i,j>} \mathbf{S}_{i} \cdot \mathbf{S}_{j} + J_{A} \sum_{<i,j>} S^{z}_{i}S^{z}_{j})}
_{\text{intra-layer part}} - \underbrace{J' \sum_{<i,k>} \mathbf{S}_{i} \cdot \mathbf{S}_{k} }_{\text{inter-layer part}}
\end{equation}
where $<i,j>$ represents the intra-layer nearest neighbour interaction and $<i,k>$ represents the inter-layer nearest neighbour interaction. In the above Hamiltonian, $J$ is the isotropic part and $J_A$ is anisotropic part of the intra-layer exchange coupling and, $J^{\prime}$ is the inter-layer exchange coupling. The ordering temperature i.e., the N$\acute{e}$el temperature for the quasi-two dimensional system ($La_2CuO_4$) is given by $T_N = 240$ K. The intra-layer part of the above Hamiltonian (\ref{eq:10}) can be simplified, by expressing $(J-J_{A})$ as $\lambda J$, to obtain the
model Hamiltonian (\ref{eq:1}), where $\lambda$ is the anisotropy parameter. The relevant physical parameters corresponding to $La_2CuO_4$ are given in the TABLE \ref{tab:Table 1} \cite{Thio}.
\begin{table}[h]
 \begin{center}
    \begin{tabular}{ | l | l | p{1cm} |}
    
    \hline
    Parameter & Magnitude\\ \hline
    J (intra-layer)&  $\sim$ 1345 K \\ \hline
    $J_A$ (intra-layer) & $\sim$ 0.269 K \\ \hline
    $J^{\prime}$ (inter-layer) & $\sim$ 0.04 K \\ \hline
    anisotropy parameter ($\lambda$) & 0.9998 \\ \hline
    lattice parameter(a)&5.39 \AA \\ \hline
    N$\acute{e}$el temperature ($T_N$) & 240 K \\ \hline
    \end{tabular}
 \end{center}
 \caption {\textit{Relevant parameters for $La_2CuO_4$}}
 \label{tab:Table 1}
\end{table}
\paragraph*{}
Next we try to determine the temperature range over which the material $La_2 CuO_4$ behaves effectively as a two-dimensional material. From the neutron scattering data for quasi-two-dimensional spin 1/2 XY-anisotropic ferromagnet $K_2CuF_4$, it was found that in a temperature regime $T_1 \leq T \leq T_2$, where the lower ($T_1$) and the upper ($T_2$) limits are defined by the following relations,
\begin{eqnarray}
\xi(T_1) &=& \sqrt{\frac{|J|}{|J^{\prime}|}} \nonumber \\
\xi(T_2) &=& \sqrt{\frac{|J|}{|J_A|}}
\end{eqnarray}\label{eq:11}
the system behaves as a 2D XY-like anisotropic system \cite{Hir, Yos}. Assuming that the above phenomenological argument holds
\begin{figure}[!h]
\centering
\resizebox{0.50\textwidth}{!}{%
  \includegraphics{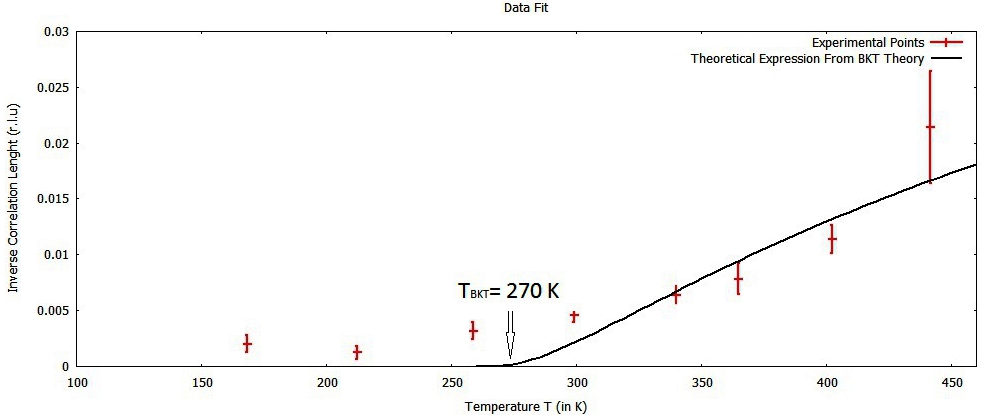}
}
\vspace{0.5cm}
\caption{\textit{Fitting of the experimentally obtained inverse correlation length with the corresponding theoretical expression (see eqn. (\ref{eq:12})). Solid line corresponds to the theoretical expression. BKT transition temperature is $T_{BKT} =270 K$.}}
\label{fig: Figure 1}
\end{figure}
for the layered anti-ferromagnetic systems as well, we determine the above two temperature limits as $T_1 \approx 260 K$ and $T_2 \approx 360 K$ for $La_2 CuO_4$. Within this temperature regime the Hamiltonian (\ref{eq:10}) can effectively be represented by the Hamiltonian (\ref{eq:1}). It is worthwhile to point out that since in the above temperature regime the system is effectively a two-dimensional one, long range anti-ferromagnetic ordering is absent in this regime \cite{Wag}. Further, within the above mentioned temperature range the BKT inspired ideal vortex/meron- gas phenomenology is valid and we can therefore use the theoretical expression for the inverse correlation length (expressed in r.l.u),
\begin{equation}\label{eq:12}
\kappa(T) = \frac{1}{\pi} e^{-b/\sqrt{\tau}}
\end{equation}
as predicted by Kosterlitz and Thouless, to fit the experimentally obtained inverse correlation length \cite{KT}. This gives the value of the BKT transition temperature as $T_{BKT} \approx 270 K$ for $La_2 CuO_4$ (see FIG. \ref{fig: Figure 1}). In this work we shall make use of the above value of $T_{BKT}$ to calculate the convoluted in-plane integrated intensity, $I_{conv}^{xx}(\omega)$ and the convoluted out-of-plane integrated intensity $I_{conv}^{zz} (\omega)$, by making use of eqns. (\ref{eq:3}) to (\ref{eq:5}), and then eqns. (\ref{eq:6}) and (\ref{eq:7}).
\paragraph*{}
We now study the convoluted in-plane integrated intensities $I^{xx}_{conv}(\omega)$ at different temperatures. The expression for the $I^{xx}_{conv}(\omega)$ is given by eqn. (\ref{eq:7}) with $\alpha = x$, where the in-plane DSF, $S^{xx}(\mathbf{q_{2D}}, \omega)$ is given by eqn. (\ref{eq:3}). In the experimental investigations on $La_2CuO_4$, to find the neutron intensity as a function of momentum transfer `$\hbar$q' the scans in the `q'-space have been performed about the zone boundary of the first BZ within the range, $-0.1 \leq \mathbf{q^*} \leq 0.1$, expressed in r.l.u \cite{Yen, Yam}. In calculating $I^{xx}_{conv}(\omega)$ using eqn. (\ref{eq:12}) we have also made use of the above mentioned regime only. The resolution function has been chosen in the form of the Tukey window to convolute the in-plane DSF. This is one of the most commonly used spectral smoothing functions in the field of spectral analysis \cite{Jen, Har}. The experimental resolution width is 1.4 meV at the full width at half maximum (FWHM) as specified in the experiment \cite{Yen, Yam}. We compute $I^{xx}_{conv}(\omega)$ for four different temperatures, viz., 290 K, 320 K, 350 K and 375 K.
\begin{figure}[!h]
\centering
\resizebox{0.45\textwidth}{!}{%
  \includegraphics{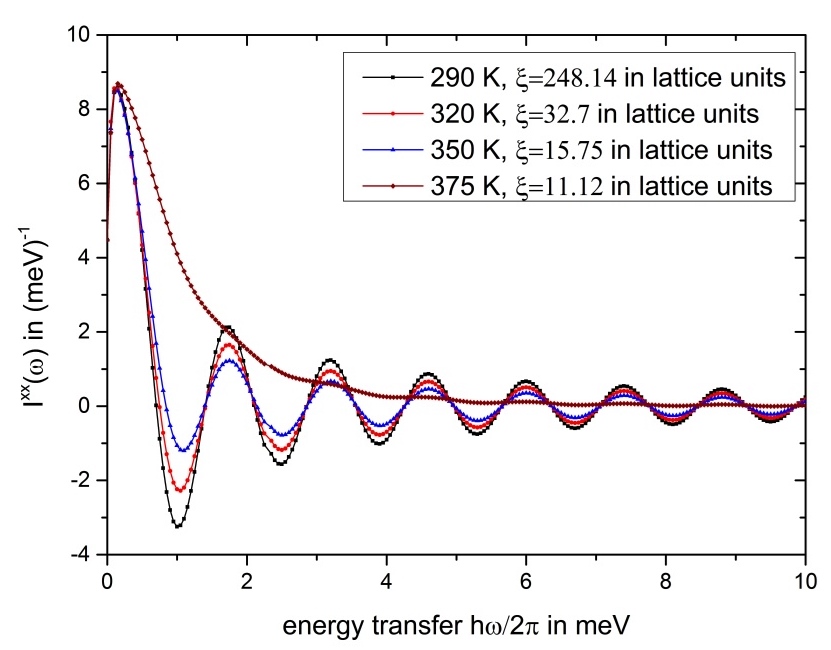}
}
\vspace{0.5cm}
\caption{\textit{The plot of convoluted (with Tukey window function) in-plane integrated intensity $I^{xx}_{conv}(\omega)|_{SC}$ at four different temperatures, viz., 290 K, 320 K, 350 K and 375 K. 
 The rms velocities at these temperatures are $\bar{u} = 0.00365 \frac{a}{t_{nat}}, \, 0.0836 \frac{a}{t_{nat}}, \, 0.085 \frac{a}{t_{nat}}, \, 0.2323 \frac{a}{t_{nat}}$ respectively.}}
\label{fig:Figure 2}
\end{figure}
The semi-classical convoluted in-plane integrated intensities denoted by, $I^{xx}_{conv}(\omega)|_{SC}$ are plotted as functions of energy transfers in FIG. \ref{fig:Figure 2}, where eqns. (\ref{eq:8}) and (\ref{eq:9}) have been used. The figures clearly exhibit that $I^{xx}_{conv}(\omega)|_{SC}$ oscillates vigorously after convoluting with the Tukey function, although the `central peak' still persists. This is quite contrary to what we experienced in the case of ferromagnet \cite{Sar}.
\paragraph*{}
To avoid such oscillations we later tried performing the above calculations using a modified version of the Tukey function (see eqns. (\ref{eq: 18}) and (\ref{eq: 19}) in the Appendix). The integrated intensity (at $290$ K) corresponding to this new window function is also plotted in FIG \ref{fig:Figure 3} along with the same corresponding to the use of Tukey function. From this figure it is clearly visible that the unwanted oscillations diminish considerably when we use the modified Tukey function. 
\begin{figure}[!h]
\centering
\resizebox{0.45\textwidth}{!}{%
  \includegraphics{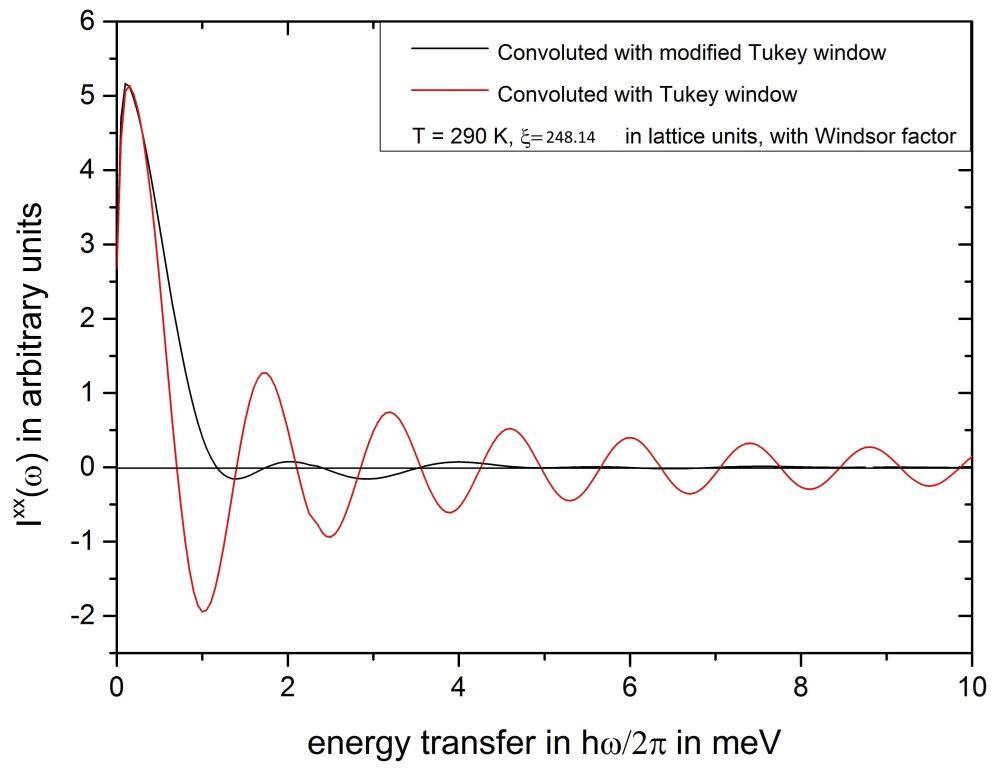}
}
\vspace{0.5cm}
\caption{\textit{The plot of convoluted in-plane integrated intensity $I^{xx}_{conv}(\omega)|_{SC}$ at 290 K. The red solid line corresponds to the $I^{xx}_{conv}(\omega)|_{SC}$ obtained by using the Tukey window function (see (\ref{eq: 18})). The black solid line corresponds to the $I^{xx}_{conv}(\omega)|_{SC}$ obtained by using the modified Tukey window function (see (\ref{eq: 19})).}}
\label{fig:Figure 3}
\end{figure} 
More interestingly, FIG. \ref{fig:Figure 4} indicates that at even higher temperature, viz., at around 375 K ($\approx 1.388 T_{BKT}$) both the Tukey function and the modified Tukey function lead to very similar results. The oscillations are totally absent in the theoretical plot of $I^{xx}_{conv}(\omega)|_{SC}$ vs. energy transfers corresponding to both the resolution functions. It is worthwhile to mention that the corresponding temperature T=375 K ($> T_2$) falls just outside the range $T_1 \leq T \leq T_2$ within which the BKT phenomenology remains valid. However, the use of such a modified Tukey function may wipe out some of the genuine and intrinsic fluctuations present in the anti-ferromagnetic systems in two dimensions. Hence, we make use of the Tukey function only for our purpose because it is very frequently used in the field of spectral analysis.
\begin{figure}[!h]
\centering
\resizebox{0.45\textwidth}{!}{%
  \includegraphics{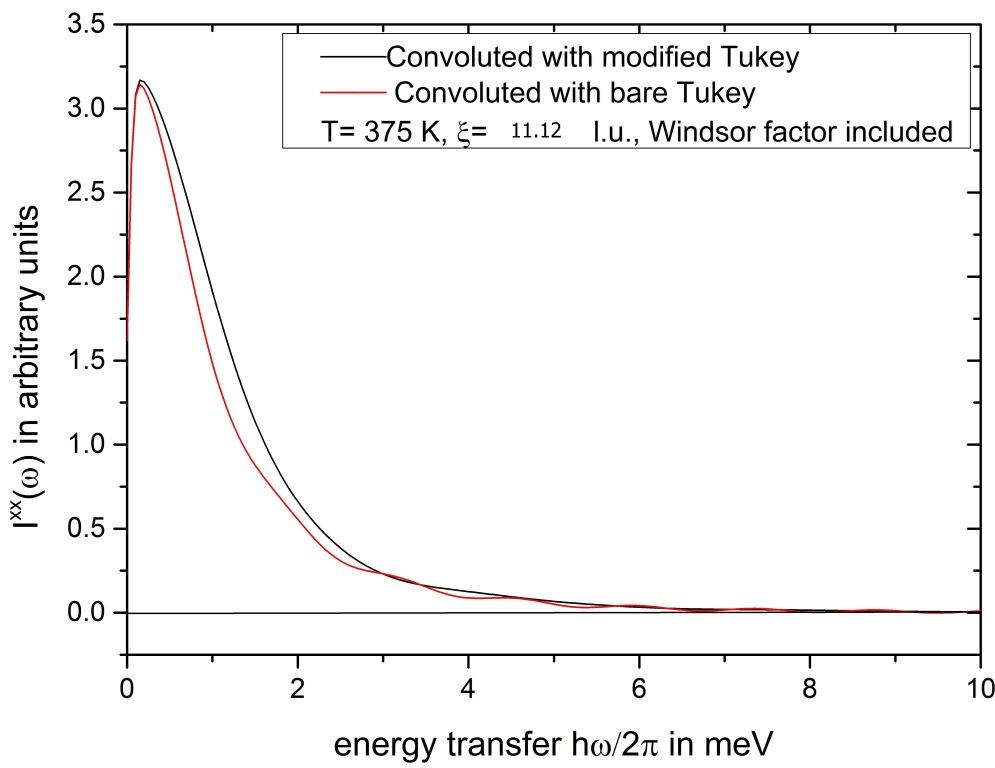}
}
\vspace{0.5cm}
\caption{\textit{The plot of convoluted in-plane integrated intensity $I^{xx}_{conv}(\omega)|_{SC}$ at 375 K. The red solid line corresponds to the $I^{xx}_{conv}(\omega)|_{SC}$ obtained by using the Tukey window function (see (\ref{eq: 18})). The black solid line corresponds to the $I^{xx}_{conv}(\omega)|_{SC}$ obtained by using the modified Tukey window function (see (\ref{eq: 19})).}}
\label{fig:Figure 4}
\end{figure} 
\paragraph*{}
We further notice a slight shift in the position of the central peak. This is due to the inclusion of detailed balance condition. The important point here is that the shift is well within the resolution width 1.4 meV at the FWHM, and hence the peak is truly a central peak situated at $\hbar \omega = 0$.
\paragraph*{}
Exactly the same results hold for $I^{yy}_{conv}(\omega)|_{SC}$ which is obvious from the symmetry argument. The normalization factor required for the quantitative comparison between the theoretical and the experimental results is estimated from the neutron count corresponding to the experimental results for $La_2 CuO_4$. 
\paragraph*{}
We now evaluate the out-of-plane integrated intensity $I^{zz}_{conv}(\omega)|_{SC}$ for the same set of temperatures as have been considered earlier for the evaluation of $I^{xx}_{conv}(\omega)|_{SC}$ (see FIG \ref{fig:Figure 7}). The expression for the $I^{xx}_{conv}(\omega)$ is given by, eqn. (\ref{eq:7}) with $\alpha = z$, where the out-of-plane DSF, $S^{zz}(\mathbf{q_{2D}}, \omega)$ is given by eqn. (\ref{eq:5}).
\begin{figure}[h]
\centering
\resizebox{0.45\textwidth}{!}{%
  \includegraphics{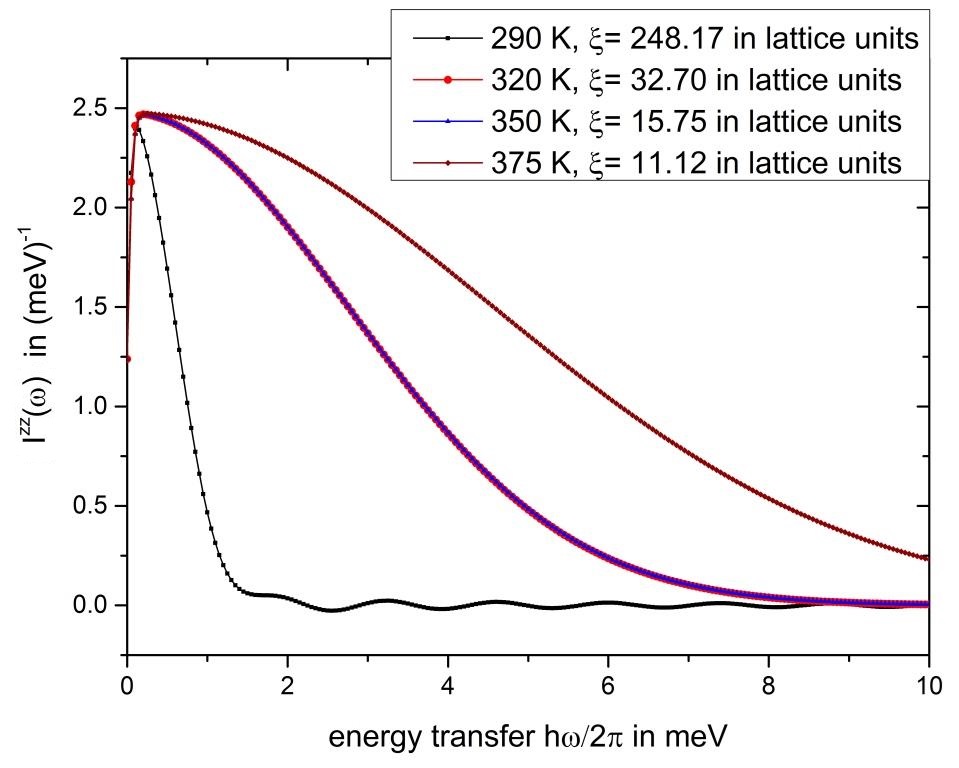}
}
\vspace{0.5cm}
\caption{\textit{The plot of convoluted out-of-plane integrated intensity $I^{zz}_{conv}(\omega)|_{SC}$ at four different temperatures, viz., 290 K, 320 K, 350 K and 375 K.
 The rms velocities at these temperatures are $\bar{u} = 0.00365 \frac{a}{t_{nat}}, \, 0.0836 \frac{a}{t_{nat}}, \, 0.085 \frac{a}{t_{nat}}, \, 0.2323 \frac{a}{t_{nat}}$ respectively. The order of magnitude of the $I^{zz}_{conv}(\omega)|_{SC}$ at the above mentioned four different temperatures are $10^{-13}$, $10^{-10}$, $10^{-8}$ and $10^{-7}$ respectively.}}
\label{fig:Figure 7}
\end{figure}
In this case we find that the out-of-plane integrated intensity oscillates only at lower temperatures near $T=T_{BKT}$. It is worth recalling here that the magnitude of the out-of-plane integrated intensity is proportional to the density $n^{f}_{v}$ and the rms velocity $\bar{u}$ of the free vortices/merons. Since both density $n^{f}_{v}$ and the rms velocity $\bar{u}$ increases with increasing temperature the out-of-plane part of the spin-spin correlation  (see eqns. (\ref{eq:5}) and (\ref{eq:7})) acquire dominance (considerable magnitude) only at higher temperatures much above $T_{BKT}$.
\paragraph*{}
Furthermore, the absolute magnitude of the integrated intensity $I^{xx}_{conv}(\omega)|_{SC}$ is higher (almost $10^7$ times for the highest temperature considered here) than that of $I^{zz}_{conv}(\omega)|_{SC}$ at temperatures above $T_{BKT}$. This is so because, $I^{zz}_{conv}(\omega)|_{SC}$ is proportional to $n^{f}_{v}$ and it increases with the increasing value of the rms velocity $\bar{u}$. The typical energy scales involved in the dynamics of mobile vortices/merons corresponding to the anti-ferromagnetic system $La_2 CuO_4$ are such that $n^{f}_{v}$ is very small (compared to the case of ferromagnet where the free vortex number density turns out to be appreciable \cite{Sar}). For the present case of $La_2 CuO_4$ at $T=350K$, the numerical value for the density of free vortices comes out to be $n^f_v = 1.36\times 10^{-4}\, a^{-2}$ and the same for the rms velocity comes out to be $\bar{u}= 9163\,  m/\sec = 0.085 \frac{a}{t_{nat}}$, where $t_{nat} = \frac{2 \hbar }{\sqrt{3} J}$ ($\approx 5 \times 10^{-15}$ sec) is the natural time scale for the system. In contrast, in the case of ferromagnetic system $K_2CuF_4$ the value of the density of free vortices was found to be $n^f_v = 1.009\times 10^{-3}\, a^{-2}$ and that for the rms velocity was found to be $\bar{u}= 87.07 \, m/\sec = 0.1352 \frac{a}{t_{nat}}$ at $T= 6.75 K$, where $t_{nat} = 6.4 \times 10^{-13} \, \sec$ is the natural time scale \cite{Sar}. 
\paragraph*{}
Interestingly enough, the unbound merons/vortices above $T_{BKT}$ move much faster (with a rms velocity of 9163 $m/\sec$ at 350 K) than a typical Copper (Cu) atom whose rms velocity (generally considered to be the thermal velocity) is around 370.6 $m/\sec$ at 350 K.
\paragraph*{}
The integrated intensities computed above correspond to the contributions only from the mobile vortices/merons. The experimental data whereas, contain contributions from both the mobile vortices/merons and fragile ``spin wave like" modes. This spin wave like modes are damped and largely decaying above the N$\acute{e}$el temperature. The extraction of the mobile vortex contribution from the experimental data is crucial for a more accurate comparison between the theoretical results and the experimental data and to do this one has to subtract the contributions from the above mentioned fragile ``spin wave like" modes from the experimental data. It is worth recalling that in the case of ferromagnetic system, the fragile mode contributions have been subtracted by assuming the fragile mode contributions above $T_c$ to be the same as the usual spin wave contributions below $T_c$. This assumption is however valid if and only if the temperature under consideration is in the near vicinity of the Curie temperature ($T_c$) of the system \cite{Sar}.
\paragraph*{}
In the present case corresponding to the anti-ferromagnetic system $La_2 CuO_4$, the situation is somewhat different from the case of ferromagnetic systems in the sense that the temperatures dealt with are far above the N$\acute{e}$el temperature $T_N$. Hence the above mentioned procedure, which was followed for the ferromagnetic systems to subtract the fragile mode contributions, is not valid.
\paragraph*{}
Moreover at any finite temperature above $T_{BKT}$, it is to be kept in mind that not all vortices/merons are freely moving and that bound vortex-anti-vortex pair density remains finite. Hence, one has to estimate further the contribution from these bound vortex-anti-vortex pairs at different temperatures and subtract them from the experimental data. To estimate the bound vortex contribution we have tried to apply the same methodology that has been outlined and used earlier for the ferromagnetic system \cite{Sar}. However, unlike the case of ferromagnet in this case the methodology leads to an unphysical behaviour viz., the vanishing of  the out-of-plane DSF (see eqn. (\ref{eq:5})) in the limit $\bar{u}\rightarrow 0$. Hence the estimation of the bound vortex contribution has not been carried out here.
\paragraph*{}
The above calculations for the components of integrated intensities enable us to estimate theoretically the total integrated intensity using (\ref{eq:8}). 
\begin{figure}[h]
\centering
\resizebox{0.45\textwidth}{!}{%
  \includegraphics{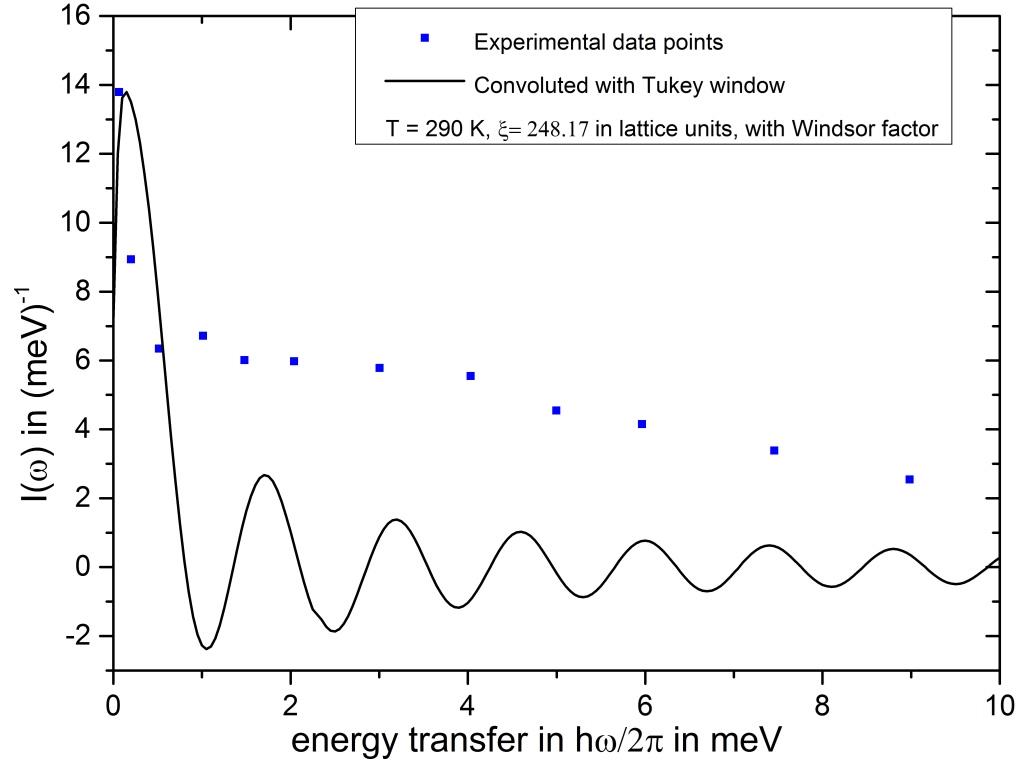}
}
\vspace{0.5cm}
\caption{\textit{The plot of convoluted total integrated intensity $I_{conv}(\omega)|_{SC}$ at 290 K. The red solid line corresponds to the $I_{conv}(\omega)|_{SC}$ obtained theoretically by using the Tukey window function  (see (\ref{eq: 18})). The dots are the experimental data.}}
\label{fig:Figure 10}
\end{figure}
\begin{figure}[h]
\centering
\resizebox{0.45\textwidth}{!}{%
  \includegraphics{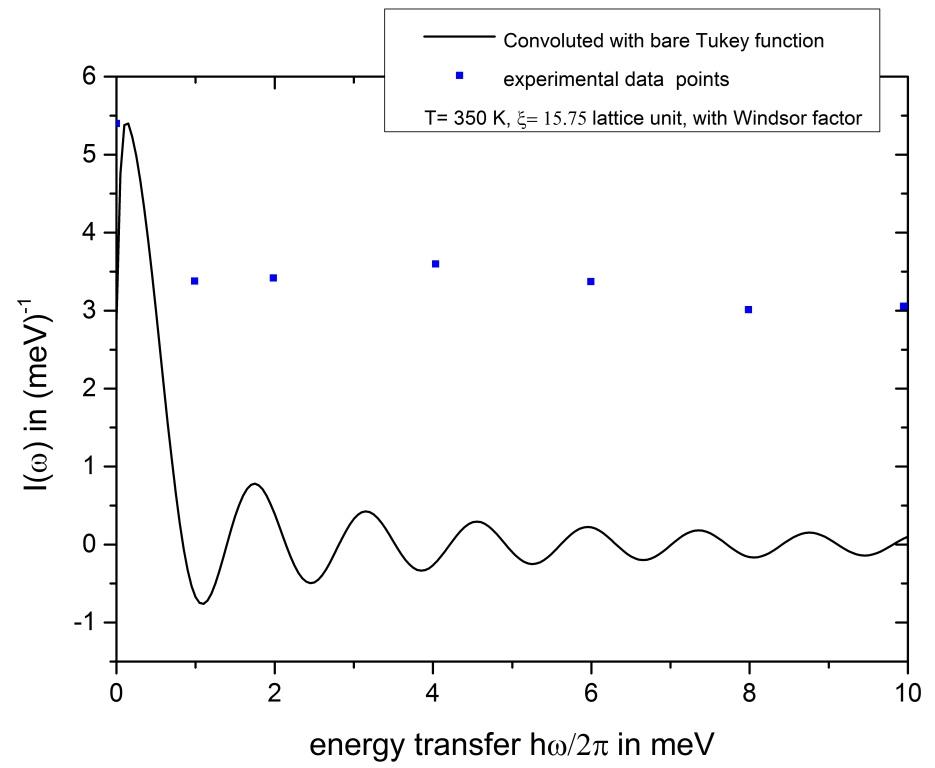}
}
\vspace{0.5cm}
\caption{\textit{The plot of convoluted total integrated intensity $I_{conv}(\omega)|_{SC}$ at 350 K. The red solid line 
corresponds to the $I_{conv}(\omega)|_{SC}$ obtained theoretically by using the Tukey window function  (see eqn. (\ref{eq: 18})). The dots are the experimental data.}}
\label{fig:Figure 11}
\end{figure}
In FIGs. \ref{fig:Figure 10} and \ref{fig:Figure 11} the total intensity $I_{conv}(\omega)|_{SC}$ has been compared to the experimental data at two different temperatures, viz., 290K and 350K respectively. The contributions from the fragile ``spin-wave-like" modes can not be filtered out for the reasons stated earlier. It is clear from FIG. \ref{fig:Figure 10} that the total intensity $I_{conv}(\omega)|_{SC}$ also oscillates vigorously at both the temperatures, when convoluted with the Tukey function. Besides, the theoretical results show negative values for $I_{conv}(\omega)|_{SC}$ with both the resolution functions! Moreover, the  magnitude of the total intensity $I_{conv}(\omega)|_{SC}$, obtained theoretically at finite energy transfers, is very far  from the corresponding values obtained in the experiment.
\paragraph*{}
The inclusion of quantum mechanical detailed balance factor in our semi-classical like treatment has again caused a shift in the position of the central peak of the integrated total intensity at both the temperatures. However, this shift is well within the resolution width and therefore it is a genuine central peak at zero energy transfer.
\paragraph*{}
It is worthwhile to mention that all the above results based on dilute vortex/meron gas phenomenology hold for unbound  anti-vortices/anti-merons too.
\paragraph*{}
Let us further calculate the zeroth moment of the semi-classical dynamical structure function (DSF) $S^{conv}_{SC}(\mathbf{q},\omega)$ (or simply the moment) using the following formula  \cite{Lov},
\begin{eqnarray}
\int_{first \, B.Z.} \, (\frac{a}{2\pi})^2 \; d^{2}\textbf{q} \int_{-\infty}^{\infty} d\omega \, S^{SC}_{conv}(\textbf{q}, \omega) &=& S(S+1), \nonumber  \\
\text{or} \;\; \int d\omega \, I_{conv}^{SC}(\omega) &=& S(S+1) \nonumber \\,
\end{eqnarray}
where $I_{conv}^{SC}(\omega)$ is given by equation (\ref{eq:9}) and in obtaining the same the integration over the wave vector space in equation (\ref{eq:7}) is performed over the first Brillouin zone (B.Z.) with contributions from both vortices and anti-vortices being summed; $S$ is the value of the spin corresponding to the system under consideration and in our case it is $S=1/2$. The above equation signifies that if the spin dynamics is entirely captured by the DSF, the value of the zeroth moment must be $S(S+1)$.
\begin{table}[H]
 \begin{center}
 \begin{tabular}{|m{3em}|m{3em}|m{8em}|}
    \hline 
    T & $\frac{T}{T_{BKT}}$ & moment in the  \\ 
     &&unit of $S(S+1)$ *\\
    \hline 
    290 K & 1.074 & 0.0400  \\ 
    \hline 
    320 K & 1.185 & 0.2412  \\ 
    \hline 
    350 K & 1.296 & 0.3583  \\ 
    \hline 
    375 K & 1.388 & 0.3770  \\ 
    \hline 
    \end{tabular}    
 \end{center}
 \caption {\textit{The zeroth moment of the semi-classical dynamical structure function corresponding to the dynamics of mobile vortices and anti-vortices}(* corresponding to the use of Tukey function.)}
 \label{tab:Table 2}
\end{table}
The values of the moment of the semi-classical convoluted DSF corresponding to the use of Tukey function at four different temperatures are tabulated in TABLE \ref{tab:Table 2}. At $290 K$ i.e. around $1.074 T_{BKT}$, the combined dynamics of mobile vortices and anti-vortices capture only about $4 \%$ of the entire spin dynamics of the system. However, at higher temperatures around $1.296 T_{BKT}$, the combined dynamics of mobile vortices and anti-vortices capture more than $35 \%$ of the entire spin dynamics of the system . This happens because at lower temperatures near $T_{BKT}$, the number of freely mobile vortices and anti-vortices are not large enough to capture the whole spin dynamics and the presence of fragile or damped spin waves (or single magnons and multi-magnon like modes) makes important enough contribution to the spin dynamics. At higher temperatures however, more topological excitations become free and drive a large portion of the spin dynamics of the system. At this point, it is worth mentioning that for quantum ferromagnetic systems on two dimensional square lattice it has been shown that the formation of topological excitations of vortex/ meron types from the fragile magnons and multi-magnon composites is quite plausible \cite{arx}. Moreover, some of the collective modes (i.e. magnon and multi-magnon modes) are expected to stay intact with their damped nature and thus can provide a significant contribution to the spin dynamics. In analogy with the three dimensional systems where above the Curie temperature ($T_c$) the magnon-like collective excitations become fragile and damped, for pure two dimensional systems (where $T_c = 0$) the collective excitations become fragile at any finite temperature \cite{Eva, Bss, Bon, Dem, Ran, Tap, arx}. This process is expected to be operative too in the anti-ferromagnetic systems on pure two-dimensional lattices.
\paragraph*{}
To summarize, we find vigorous oscillations in the convoluted in-plane integrated intensity when the Tukey window is used. These oscillations vanish only at higher temperatures which are outside the regime of validity of the BKT phenomenology. The use of a modified or refined Tukey function substantially removes the unwanted oscillations in the convoluted in-plane integrated intensity $I^{xx}_{conv}(\omega)|_{SC}$. Strikingly enough, at T = 350 K (1.296 $T_{BKT}$), we still find negative values of $I^{xx}_{conv}(\omega)|_{SC}$ even using the modified Tukey window function. However, outside the temperature regime where the BKT phenomenology is valid, computations with both the window functions give very similar results for $I^{xx}_{conv}(\omega)|_{SC}$. The possible explanation for this is that at higher temperatures the quantum effects are less prominent even for $S= \frac{1}{2}$ anti-ferromagnet. Therefore, the modified Tukey window may actually be suppressing quantum fluctuation as well quite efficiently.
The out-of-plane integrated intensities $I^{zz}_{conv}(\omega)|_{SC}$  (computed at different temperatures) are found to be sensitive to the choice of window function only at temperatures which are not very far from $T_{BKT}$ (around 1.074 $T_{BKT}$). Furthermore, it contributes negligibly to the total integrated intensity $I_{conv}(\omega)|_{SC}$ and hence the nature of the convoluted total integrated intensities at different temperatures turns out to be quite similar to that of the convoluted in-plane integrated intensities. However, the detailed quantitative comparison between the theoretical results and the experimental results corresponding to the total integrated intensities $I_{conv}(\omega)|_{SC}$ reveals that even though our ``semi-classical like" theory is able to predict the occurrence of the central peak, the magnitudes of the $I_{conv}(\omega)|_{SC}$ for finite values of energy transfer, obtained from theoretical analysis, differ by a huge factor from the corresponding experimental values. Moreover, apart from the spin dynamics induced by the mobile topological excitations, the fragile magnons and multi-magnon modes are quite likely to make important contribution towards this dynamics.
\section{Conclusions and discussions}
It is a well known fundamental fact that the spin-dynamical structure function (DSF) and hence the integrated intensity corresponding to all real magnetic systems must be positive definite \cite{Lov, Gen, Hip}. Our detailed calculations and analysis however, brings out an important fact that for  quasi-two dimensional low-spin anti-ferromagnetic systems, the semi-classical treatment of ideal gas of mobile vortices/merons (anti-vortices/anti-merons) leads to negative values of total integrated intensities even in the low energy regime, when convoluted with commonly used resolution functions. Similar results were obtained for $S=1/2$ layered ferromagnets \cite{Sar}. Such a behaviour is purely unphysical. These facts highlight the inapplicability of the semi-classical vortex/meron gas phenomenology to the quasi-two dimensional (or rather layered) low-spin anti-ferromagnetic systems very strongly and necessitates the demand for a full quantum treatment of the problem. Let us further point out that the value of $\hbar \omega$ at which the onset of such unphysical behaviour occurs depends on the value of the spin $(S)$. It has been shown in the case of ferromagnetic systems that the regime over which this unphysical behaviour persists, shrinks as the value of $S$ increases \cite{Sar}. Similar behaviour is expected in the anti-ferromagnetic systems also. This may be verified from the studies on the compounds based on Gd and Mn. \\

\paragraph*{}
To the best of our knowledge there are no materials based on $Gd^{3+}$ (spin $\frac{7}{2}$) which can be modelled by 2D XXZ Hamiltonian (XY anisotropic Heisenberg Hamiltonian). However, some high spin magnetic systems such as $Rb_2 MnF_4 , \, Cs_2 Mn Cl_4$, both being spin $\frac{5}{2}$ compound, turn out to be a layered Heisenberg AFM with Ising like anisotropy \cite{Thu}. In this case the BKT scenario does not hold. On the other hand applicability of the Berezinskii-Kosterlitz-Thouless scenario on triangular chromium-lattice AFM with $S=3/2$ has been investigated only via electron spin resonance (ESR) technique \cite{Hem}. In this regard it is worthwhile to mention that the spin dynamics in a $Mn^{2+}$ based spin $\frac{5}{2}$ Honeycomb lattice anti-ferromagnetic material $MnPS_3$ has been investigated via INS experiment \cite{Wild1}. The critical properties of this material have been reported to be well described by 2D XXZ Hamiltonian (anisotropy parameter being equal to 0.998) only in the low $\mathbf{q^*}$ regime. We have performed an initial study towards the calculation of DSF for $MnPS_3$ using our model and this reveals that the semi-classical BKT phenomenology is producing a much better agreement with the experimental data corresponding to this material at 85 K \cite{Wild2}. The agreement is only in terms of peak position and the range of energy transfer over which we get acceptable (positive) values of DSF. This range encompasses a considerable portion of the range of experimental interest. Regardless of this, the possibility of meron dynamics in this material can not be adequately described with the available INS data. To be very specific, the available INS data is at 85 K whereas, the critical dynamics is completely confined within the plane (and therefore, corresponding to a perfect 2D system) only above 105 K \cite{Wild1}. Furthermore, in this material $T_N > T_{BKT}$ and therefore, above $T_{BKT}$ and below $T_N$ well defined magnon modes persist which is in contrast to the case corresponding to $La_2 CuO_4$.
\paragraph*{}
 Presently it seems that there is a scarcity of INS data in search for the spin dynamics induced by topological excitations corresponding to XY anisotropic (easy plane anisotropic, i.e., XXZ type) layered Heisenberg anti-ferromagnetic materials having higher values of spin. More INS experiments on this type of materials (in a suitable temperature range) would be quite interesting in view of the fact that it would serve as a good testing ground for the applicability of such a conventional semi-classical theory of spin dynamics induced by topological excitations.
\paragraph*{}
It is also very important to emphasize the fact that in our calculational analysis the structure of the classical vortex/meron has been built in the background of the N$\acute{e}$el state (see (\ref{eq:2}) of section II). Since the N$\acute{e}$el state is not an exact ground state for the two-dimensional quantum anti-ferromagnetic spin systems, such a choice further adds to the reasons for the above mentioned unphysical behaviour.
\paragraph*{}
Keeping aside the occurrences of unphysical negative values of the integrated intensities obtained from the conventional semi-classical-like theory, the results agree with those from the experiment quite well in terms of the existence of the central peak. However, this peak is a genuine one only in the sense that it occurs well within the experimental resolution width.
\paragraph*{}
 Moreover, quantitative disagreement, at the finite values of energy transfer, between the magnitudes of the theoretically calculated integrated intensities and those obtained experimentally strongly indicates that formation of the topological excitations from fragile collective modes and interactions between them can play a very crucial role in the spin dynamics. The values of the moment at different temperatures further strongly suggest the same. Therefore, the interplay between the fragile single magnons or multi-magnon composites and the topological excitation is very crucial for the proper understanding of the dynamics induced by free movement of the latter in the quasi-two-dimensional low-spin anti-ferromagnetic systems. This further takes care of the fragile magnon contributions to the total integrated intensities as well which has not been filtered out from the  experimental data (for the reasons stated earlier in the section III). In addition, the severe inadequacy of the semi-classical treatment in the case of quantum anti-ferromagnet may also contribute to such quantitative disagreement for the integrated intensities.
\paragraph*{}
Our investigations presented in this communication establish the fact that a complete quantum treatment is essential to describe the detailed features of the dynamics of mobile topological excitations corresponding to the quasi-two-dimensional low-spin anti-ferromagnetic systems, by taking into consideration the vortex/meron-fragile magnon interactions as well. This further brings out the possibility of a characterization of different layered materials into two classes, viz., conventional BKT-like and quantum corrected BKT-like. Moreover, since the N$\acute{e}$el state is not an exact ground state for the two-dimensional quantum spin systems, the construction of the quantum vortices/merons in the complete quantum treatment has to be supplemented with proper choice of the ground state too. However, calculation of the dynamical structure function in a completely quantum mechanical framework is highly non-trivial. Earlier attempts towards this goal could not explain the occurrence of the ``central peak" in the DSF obtained in the INS experiment performed on several quasi-two-dimensional materials \cite{Yam, CHN, AA, DeJ}. Quite recently a theoretical framework had been developed based on the spin coherent state path integral formalism to describe the topological properties of static vortices and anti-vortices \cite{Hal, Fra, Fdm, Skp, Skp1, Skp2}. An extension of this framework to the case of mobile spin vortices/merons (and anti-vortices/anti-merons) is crucial for quantum mechanical calculations of the DSF and the integrated intensity as well. Insights from Quantum Monte Carlo calculations may be quite useful in this endeavour \cite{Vit}. This would be taken up in future. More importantly the short range 2D anti-ferromagnetic spin-spin correlation persists even in the superconducting phase of the underdoped cuprates. This has been investigated and verified via INS experiments \cite{Kast}. Moreover it has been shown very recently that the spin fluctuations in the AFM quantum critical region of the Fe based superconductors and some heavy fermion compounds can be modelled by dissipative quantum XY model and hence, the static and dynamics of topological excitations are the key factors for 2D spin-spin correlations \cite{CMV1, CMV2}. Therefore, this investigation of the dynamics of the BKT vortices/merons may also contribute substantially to the microscopic understanding of lightly doped anti-ferromagnetic cuprates and the above mentioned other systems as well \cite{RC, Chen, CMV1, CMV2}.
\subsection*{Acknowledgements:}
One of the authors (SS) acknowledges the financial support through Senior Research Fellowship (09/575 (0089)/2010 EMR–1) provided by Council of Scientific and Industrial Research (CSIR), Govt. of India. 
\subsection*{Author contribution statement:}
All authors have contributed equally to this communication.
\appendix
\section{ The Tukey and the modified Tukey function}
The most general form for the Tukey function is given by,
\begin{widetext}
\begin{equation} \label{eq: 16}
    R(t)= 
\begin{dcases}
    \frac{1}{2} [1+cos(\frac{\pi}{1-\alpha} \frac{2t}{t_m} + \frac{\pi}{1-\alpha} - \pi)],& \text{for } \frac{-t_m }{2} \leq t < \frac{-t_m }{2} \alpha \\
    1,              & \text{for } \frac{-t_m }{2}\alpha  \leq t \leq \frac{t_m }{2} \alpha \\
    \frac{1}{2} [1+cos(\frac{\pi}{1-\alpha} \frac{2t}{t_m} - \frac{\pi}{1-\alpha} + \pi)],& \text{for } \frac{t_m }{2}\alpha < t \leq \frac{t_m }{2} \\
    0, & \text{otherwise},
\end{dcases}
\end{equation}
\end{widetext}
where $\alpha$ is called the tapering parameter \cite{Jen, Har}. 
\paragraph*{}
The Tukey function we have used in this communication is corresponding to $\alpha = 0$. In this case the above general 
expression takes the form,
The Tukey function is given by,
\begin{equation} \label{eq: 17}
    R(t)= 
\begin{dcases}
    \frac{1}{2} [1+cos(\frac{2 \pi t}{t_m})],& \text{for } |t| \leq \frac{t_m}{2}  \\
    0, & \text{otherwise},
\end{dcases}
\end{equation}
which is corresponding to zero tapering \cite{Jen, Har}. The Fourier transform of the above Tukey function (corresponding to zero tapering) 
is given by,
\begin{equation}
\tilde{R}(\omega)= \frac{1}{4 \pi} sin (\frac{\omega t_m}{2}) ( \frac{2}{\omega} - \frac{1 }{\omega + \frac{2 \pi}{t_m}} - \frac{1}{\omega - \frac{2 \pi}{t_m}} ).
\end{equation}
The full width at half maximum (FWHM) for the above function is given by ${\Delta}_{FWHM}^{(T)} = \frac{4\pi}{t_m}$, 
expressed in the units of energy, where the superscript T signifies the Tukey function \cite{Bss}. To find the value of $t_m
$ the above expression for $\Delta_{FWHM}^{(T)}$ is equated to the value of the resolution width, 1.4 meV at FWHM, specified 
in the experimental observations corresponding to $La_2 CuO_4$.
\paragraph*{}
The modified Tukey (MT) function we have used in our analysis is corresponding to 50\% tapering and it is given by \cite{Jen, Har},
\begin{equation} \label{eq: 18}
    \mathcal{R}(t)= 
\begin{dcases}
    \frac{1}{2} [1- cos(\frac{4 \pi t}{t_m})],& \text{for } \frac{-t_m }{2} \leq t \leq \frac{-t_m }{4} \\
    1,              & \text{for } \frac{-t_m }{4} \leq t \leq \frac{t_m }{4} \\
    \frac{1}{2}[1- cos(\frac{4 \pi t}{t_m})],& \text{for } \frac{t_m }{4} \leq t \leq \frac{t_m }{2}\\
    0, & \text{otherwise}.
\end{dcases}
\end{equation}
Fourier transform of the above modified Tukey function is given by,
\begin{widetext}
\begin{equation} \label{eq: 19}
\tilde{\mathcal{R}}(\omega)= \frac{1}{4 \pi} [sin (\frac{\omega t_m}{2}) + sin(\frac{\omega t_m}{4})]  ( \frac{2}{\omega} - \frac{1 }{\omega + \frac{4 \pi}{t_m}} - \frac{1}{\omega - \frac{4 \pi}{t_m}} ).
\end{equation}
\end{widetext}
The full width at half maximum corresponding to the above function $\tilde{\mathcal{R}}(\omega)$ can be found out to be $
\Delta^{(MT)}_{FWHM} \approx \frac{3.2 \pi}{t_m}$, expressed in the units of energy. In this case, to find the value of $t_m
$ we have equated $\Delta_{FWHM}^{(MT)}$ to the value of the resolution width specified in the experiment.

\end{document}